\documentclass{hep99t}
\begin{document}
%
\thispagestyle{empty}
\onecolumn
\begin{flushright}
\large
YITP-99-55 \\
October 1999
\end{flushright}
\vspace{1.2cm}

\renewcommand{\thefootnote}{\fnsymbol{footnote}}
\setcounter{footnote}{1}
\begin{center}  
\begin{LARGE} 
\begin{bf}
Resummation and power corrections \\
for factorized cross 
sections\footnote{
\begin{large} 
Talk presented at the ``International 
Europhysics Conference on High Energy Physics'', 
15-21 July 1999, Tampere, Finland
\end{large}}

\end{bf}  
\end{LARGE}

\vspace*{1.8cm}
{\Large George~Sterman and  Werner~Vogelsang}

\vspace*{4mm}

\begin{large}
{C.N.\ Yang Institute for Theoretical Physics, State University of
New York, Stony Brook}

\vspace*{2mm}
Stony Brook, NY 11794-3840, U.S.A.\\[3pt]
\vspace*{2mm}
E-mails: {\tt sterman@insti.physics.sunysb.edu,
vogelsan@insti.physics.sunysb.edu}
\end{large}
\vspace*{2.cm}

%

{\Large \bf Abstract}
\end{center}

\noindent
Threshold resummation for
factorizable cross sections in
hadron-hadron collisions has a number of
applications and extensions.  We discuss factorization scale
dependence,
resummation at nonleading power in the moment variable,
and the implications of resummation for power corrections
in the hard momentum scale.

\setcounter{page}{0}
\renewcommand{\thefootnote}{\arabic{footnote}}
\setcounter{footnote}{0}
\normalsize
\newpage

\def \cl {{\cal L}}
\def \d {{\rm d}}
\def \be {\begin{equation}}
\def \ee {\end{equation}}
\def \bea {\begin{eqnarray}}
\def \eea {\end{eqnarray}}
\def \bi {\bibitem}
\def \ci {\cite}
\def \e {{\rm e}}
\def \o {\omega}
\def \a {\alpha}
\def \n {\nu}
\def \cf {{\cal F}}
\def \x {\xi}
\def \g {\gamma}
\def \del {\partial}
\def \pt {{_{\rm PT}}}
\def \eps {\varepsilon}
\newcommand\re[1]{(\ref{#1})}
\def \lab #1 {\label{#1}}

\def \p {\pi}
\def \m {\mu}
\def \cs {{\cal S}}
\def \as {{\alpha_s}}
\def \CO {{\cal O}}
\def \msb {{\rm(\overline{MS})}}
\def \to {\rightarrow}

\title{Resummation and power corrections for factorized cross sections}

\author{George Sterman and Werner Vogelsang}
%
%

\address{C.N.\ Yang Institute for Theoretical Physics, State University of
New York, Stony Brook\\
Stony Brook, NY 11794-3840 U.S.A.\\[3pt]
E-mails: {\tt sterman@insti.physics.sunysb.edu,
vogelsan@insti.physics.sunysb.edu}}

\abstract{Threshold resummation for
factorizable cross sections in
hadron-hadron collisions has a number of
applications and extensions.  We discuss factorization scale
dependence,
resummation at nonleading power in the moment variable,
and the implications of resummation for power corrections
in the hard momentum scale.
}

\maketitle

\section{Threshold corrections}

The factorization theorem for hard
 cross sections in QCD may be represented as
\bea
\sigma_{AB\to F}(Q)
&=& \phi_{a/A}(x_a,\mu) \otimes
\phi_{b/B}(x_b,\mu)   \nonumber \\
&\ & \hspace{-25 mm}\otimes\;
\hat{\sigma}_{ab\to F}\left(z\equiv
Q^2/x_ax_bS,Q,\mu,\alpha_s(\mu^2)\right)\, ,
\label{fact}
\eea
with the $\phi$'s parton distributions and $\hat \sigma_{ab\to F}$
partonic cross sections, calculable in perturbation theory.
The point $z=1$ corresponds to ``partonic threshold", at which
the partons $a$ and $b$ have just enough energy to produce
the observed final state, $F$, and at which $\hat \sigma$ is
generally singular.  These singularities may be illustrated
by moments of the inclusive Drell-Yan cross
section,\bea
\int_0^1 d\tau \tau^{N-1} {1\over \sigma_0}
{d\sigma_{AB\rightarrow V} \over d Q^2}(\tau\equiv Q^2/S, Q^2)
 &=&
\nonumber\\
&\ &
\hspace{-55 mm} \sum_{a,b=q{\bar q},g}
{\tilde \phi}_{a/A}(N,\mu) \,
{\tilde \phi}_{b/B}(N,\mu) \nonumber \\
 &\ & \hspace{-50 mm} \times\; {\tilde \omega}^{\rm DY}_{ab}
\left(N, Q,\mu,\alpha_s(\mu^2)\right)\, .
\label{moment}
\eea
In this case, the singular corrections in $z$ and their
finite, but logarithmic, moments are of the form
\be
\left [ {\ln^{2k-1}(1-z)\over 1-z} \right ]_+
 \to \ln^{2k}N\, ,
\nonumber
\ee
at $k$th order in $\alpha_s$.  The term {\em threshold resummation}
refers to the summation of all such singular corrections to all orders
in $\alpha_s$.
Such a resummed form has been known for some time for the Drell-Yan
cross section \cite{oldDY,ct}, where the hard-scattering function
$\omega(N)$ is given by an exponential,
in $\overline{\rm MS}$ scheme as
\bea
{\tilde \omega}^{\rm DY}_{ab}
(N)&\sim&\delta_{b{\bar a}}\;  \exp \Bigg [-\int^1_0 dz \frac{z^{N-1}-1}{1-z}
\nonumber \\ && \hspace{-20mm}
\times\;  \int^{\mu^2}_{(1-z)^2Q^2} \frac{d\xi^2}{\xi^2}
A^{(a{\bar a})}[\alpha_s(\xi^2)]\Bigg]\, ,
\label{omegaresum}
\eea
where $A$ is a finite function of the running coupling,
known explicitly to two  loops,
\bea
A^{(ab)} &=& (C_a+C_b)\left ( {\alpha_s\over \pi}
+\frac{1}{2} K \left({\alpha_s\over \pi}\right)^2\right )
\nonumber\\
K&=& C_A\; \left ( {67\over 18}-{\pi^2\over 6 }\right ) - {5\over 9}n_f\, ,
\eea
with $C_q\equiv C_F,\ C_g\equiv C_A$
In this form, all logarithms of $N$, and hence singular distributions
in $1-z$, arise from the explicit integrals in Eq.\ (\ref{omegaresum}),
with the coupling running as shown.

Inverting expressions such as (\ref{omegaresum}) gives
estimates of the corresponding cross section that
include corrections from all orders. \cite{invert}
Normally, the results do not drastically modify
next-to-leading order predictions.  At
the same time, threshold-resummed cross sections
have a much reduced dependence on the factorization
scale $\mu$ in Eq.\ (\ref{fact}). \cite{cmnt,scale}  Yet another
application is to study the
strong coupling limit $\xi^2 \rightarrow0$ in
Eq.\ (\ref{omegaresum}), as a guide to
power corrections.

\section{Refactorization and scale dependence}

The reduction of scale dependence in resummed cross sections
may be understood by ``refactorizing" the partonic
cross section in terms of parton distributions $\psi_{a/h}(x)$ defined
for a hadron (or parton) $h$, of momentum $p$,
at measured partonic ($a$) {\it energy} $xp_0$, rather than fractional
momentum $xp^\mu$, \cite{oldDY}
\bea
\int_0^1 d\tau \tau^{N-1} {1\over\sigma_0}
{d\sigma_{ab\rightarrow V} \over d Q^2}(\tau, Q^2) &\ &
\nonumber\\
&\ & \hspace{-45mm}= {\tilde \phi}_{a/a}(N,\mu) \,
\, {\tilde \omega}^{\rm DY}_{ab}
\left(N, Q,\mu\right) \bullet \nonumber\\
&\ & \hspace{-45mm}
= {\tilde\psi}_{a/a}\left (N,Q\right )\;
U_{ab} ( N) \bigg | H_{ab} \left({Q}\right)\bigg|^2
\bullet\, ,
\label{phipsifact}
\eea
where we have exhibited the distribution for incoming
parton $a$, with the corresponding factors for
parton $b$ denoted by ``$\bullet$".
The flavor-diagonal energy distributions $\psi_{a/a}(x)$ match the phase space
of the annihilation process to ${\cal O}(1/N)$, and
automatically absorb all leading logs
in $N$ near threshold.  Most importantly, they are ultraviolet
finite, and are thus {\it independent of the factorization
scale $\mu$}.
The remaining logs are
coherent in the incoming partons $a$ and $b$, and
are absorbed into an ``eikonal"
function $U_{ab}(N)$.

Equating the factorized and refactorized partonic cross sections
in Eq.\ (\ref{phipsifact}), we readily derive
\be
{\tilde \omega}^{\rm DY}_{ab}(N,Q,\mu)
=
\left[ {{\tilde\psi}_{a/a}(N,Q) \over {\tilde\phi}_{a/a}(N,\mu)}\right]^2
U_{ab}
|H_{ab}|^2\bullet\, ,
\label{tildeo}
\ee
in which the dependence on $\mu$ appears only in the
denominators ${\tilde \phi}(N,\mu)$.
 The ratios of ${\tilde\psi}$ to ${\tilde \phi}$ may be
computed, to exhibit the exponentiation of $N$-dependence
in Eq.\ (\ref{omegaresum}) above.

Inserting (\ref{tildeo}) into the hadronic cross
section,  we have, in the same notation,
\bea
\sigma^{\rm DY}_{AB}(N)  &=&
{\tilde \phi}_{a/A}(N,\mu)
{\tilde \omega}^{\rm DY}_{ab}(N,Q,\mu)\;
\bullet\\
&\ & \hspace{-15mm}
=
{\tilde \phi}_{a/A}(N,\mu)
\left[ {{\tilde\psi}_{a/a}(N,Q) \over {\tilde\phi}_{a/a}(N,\mu)}\right]^2
U_{ab}
|H_{ab}|^2\;
\bullet\, .
\nonumber
\eea
Noting, however, that the scale dependence of the ${\tilde \phi}$'s is {\em
universal},
we have
\be
\mu{d\over d\mu}\left [ {{\tilde \phi}_{a/A}(N,\mu)\over {\tilde
\phi}_{a/a}(N,\mu)} \right]=0\, ,
\ee
so that $\mu$-dependence disappears (to ${\cal O}(1/N)$), \cite{cmnt}
as observed in explicit applications of the formalism.

\section{Resummation beyond logs of $N$}

The resummation described above organizes all logs
of the moment variable $N$.
At this, leading, power in $N$ we may neglect mixing
between parton species.  We must, however,
recall the long-known
result that at high evolution scales, the large-$N$ behavior of the
gluon distribution is eventually dominated by the quarks,
\be
G(N,\mu) \sim {\Sigma(N,\mu) \over N \ln N}\, ,
\ee
where $\Sigma$ is a singlet combination of quark distributions.
There is thus a need to extend the resummation formalism to ${\cal O}(1/N)$.

Contributions of the form $\ln^m N/N$, in particular,
are potentially resummable, since they correspond to
integrable, but logarithmically divergent behavior
in the $x\to 1$ limit.
The phenomenological importance of these contributions has been
stressed for Higgs production cross sections. \cite{kls}
So far, resummation at order $1/N$ has been
carried out only in the case  of moments of $F_L(x,Q)$ in
DIS, which begin at that power.  \cite{as}

The extension to
$F_2$ and hadron collisions requires an analysis that
generalizes Eq.\ (\ref{tildeo}), by taking parton
mixing into account,
\bea
{\tilde \omega}^{\rm DY}_{ab}(N,Q,\mu)
&=& \nonumber\\
&\ & \hspace{-26mm}
{\tilde\phi}_{e/a}(N,\mu)}^{-1}\, {{\tilde\psi}_{c/e}(N,Q)
\; \rho_{cd}(N,Q)\;
\bullet\, ,
\eea
where the ${\tilde\psi}$ and ${\tilde \phi}$ are now matrices in
flavor, with off-diagonal
elements beginning at ${\cal O}(1/N)$. Corrections to this result are
suppressed by
 powers of $Q$ rather than $N$, again because
of the universality of mass factorization.  The new
``hard-scattering functions" $\rho_{cd}(N,Q)$ will
in general include corrections of all powers in $1/N$,
although, as before, the factorization-scale dependence
is contained in the ratios of ${\tilde\psi}$'s to ${\tilde \phi}$'s.
These ratios have a form  reminiscent of NLO parton
distribution evolution in the singlet sector,
\bea
{\tilde\phi}^{\msb}_{e/a}{}^{-1}(N,Q)}\; {{\tilde\psi}_{c/e}(N,Q)  &=&
\bar{{\cal P}} \exp \Bigg \{
\int_{Q^2\over N^2}^{Q^2} \frac{d\xi^2}{\xi^2}
\nonumber\\
&\ & \hspace{-42mm} \times
\left[ \bar{\Gamma} (N,\xi^2) - A(\alpha_s(\xi^2)) \ln (\frac{N^2
\xi^2}{Q^2} ) \right] \Bigg \}_{ca}\, ,
\eea
as path-order exponentials in terms of a matrix of anomalous dimensions,
$\Gamma_{ab}$,
which is
entirely of order $1/N$, and a diagonal matrix $A_{ab}$.

\section{Power corrections from resummation}

As is evident in Eq.\ (\ref{omegaresum}), resummed cross sections
involve integrals for which the argument of the  running
coupling vanishes.  Over the past few years, it has been
realized that these results give a way to predict the form
of power corrections to cross sections to which the operator
product expansion does not directly apply. \cite{cost,irr,bb}

\noindent
In $\rm e^+e^-$ annihilation, this approach
has been employed to organize power corrections to event shapes.
For the thrust, the
analysis of resummation formulas similar to
(\ref{omegaresum}) leads to a convolution form
for the cross section that organizes all terms of
order ${\cal O}([N/Q]^a)$ in moment space: \cite{ks99}
\be
{d\sigma (t) \over
dt} =
\int_0^{tQ} d\epsilon\; f(\epsilon)\; {d\sigma_{\rm PT}(t-\epsilon/Q) \over
dt} +{\cal O}\left({1\over tQ^2}\right)\, ,
\ee
with $f(\epsilon)$ a new  nonperturbative function.  The well-known
prescription of a nonperturbative ``shift" ($t\to t-\lambda_1/Q$)
in the resummed thrust distribution
\cite{ksmordw} follows in the limit of a
very narrow $f(\epsilon)\sim \delta(\epsilon-\lambda_1)$.

A related question, which has seen less attention,
is the nature of power
corrections in hadronic inclusive cross sections, such as
Drell-Yan. \cite{cmnt} In particular, it was suggested  in Ref.\
\cite{cost}
on the basis of Eq.\ (\ref{omegaresum})
that all integer powers $(N/Q)^a$ occur in Drell-Yan.
Subsequently, however, it was shown that the coefficient of the
linear,  $N/Q$, term in this series vanishes,
when the anomalous dimension $A$ is
calculated to all orders in perturbation
theory in the  large-$n_f$ limit \cite{bb}, a procedure not practical
in full QCD.

We have returned to this issue recently, verifying that
the $N/Q$ correction is indeed absent in full QCD.  We
can see this on the basis of rather general arguments, without
recourse to explicit all-orders calculations, although we must generalize
the form of Eq.\ (\ref{omegaresum}) somewhat.

The basic observation is illustrated by the following
expression for the ratio of ${\tilde\psi}$ to ${\tilde \phi}$ functions,
relevant to resummed Drell-Yan, given here at lowest order in $A$:
\begin{eqnarray}
\ln\left[{{\tilde\psi}_{i/i}(N,Q)\over {\tilde \phi}_{i/i}^{\msb} (N,Q)}\right]
&\sim&
\int_0^1 {dz\over 1-z}\int_0^{2(1-z)^2Q^2}{dk_T^2\over k_T^2} \nonumber\\
&\ & \hspace{-32mm} \times
\left[ {\rm e}^{-N[(1-z)+k_T^2/2(1-z)Q^2]} -{\rm e}^{-N(1-z)}
\right]{\alpha_s\over 2\pi}
\nonumber\\
&\ & \hspace{-32mm} - \int_0^1 {dz\over 1-z}\left({\rm e}^{-N(1-z)}-1\right)\;
\int_{2(1-z)^2Q^2}^{Q^2}{dk_T^2\over k_T^2} {\alpha_s\over 2\pi}\, .
\nonumber\\
\label{fullalpha}
\end{eqnarray}
The first integral on the right-hand side illustrates the finite remainder
after
the cancellation  of collinear divergences between the energy (${\tilde\psi}$)
and $\overline{\rm MS}$ (${\tilde \phi}$) distributions.   It has  no logs
of $N$.  This term is explicitly present in
the analysis of Ref.\ \cite{oldDY},
but was treated only to lowest order, and absorbed into an overall constant
in the
resummed cross section, Eq.\ (\ref{omegaresum}).
To study power corrections due to  the full ${\cal O}(N^0)$
cross section, however, it is necessary to consider the running of
the coupling in this contribution. The second integral on the right-hand
side contributes
directly to the resummed cross section of Eq.\ (\ref{omegaresum}), and
contains  all logs of $N$.

Power corrections in the ratio of ${\tilde\psi}/{\tilde \phi}$ can be
identified
by expanding the $z$ integrals of (\ref{fullalpha}) in powers of $k_T/Q$
in the strong-coupling ($k_T\to 0$) limit, running the coupling with scale
$k_T$. \cite{cost,irr,bb}  In Eq.\ (\ref{fullalpha}), we verify that
the  $1/Q$ terms found in this way cancel between the first and second
integrals.
Power corrections in the ratio, and in the full cross section, thus begin
at $(N/Q)^2$.
This analysis can be extended to all orders in perturbation theory.

\section*{Acknowledgments}
We thank Stefano Catani, Nikolaos Kidonakis, Michelangelo Mangano, Paolo Nason
and Jeff Owens for many helpful conversations on resummation.
 This work was supported in part by the National Science Foundation, grant
PHY9722101.

\end{document}